\documentclass[letter]{aa}  

\usepackage{natbib}
\bibpunct{(}{)}{;}{a}{}{,} 

\usepackage{graphicx}
\usepackage[varg]{txfonts}
\newcommand{\kms}{km\,s$^{-1}$}

\begin{document}

\title{Detection of a new methanol maser line with ALMA}
\author{I. Zinchenko\inst{1} \and
S.-Y. Liu\inst{2} \and
Y.-N. Su\inst{2} \and
A.~M.~Sobolev\inst{3}
}
\institute{Institute of Applied Physics of the Russian Academy of Sciences,
	   46 Uljanov~str., 603950 Nizhny Novgorod, Russia\\
              \email{zin@appl.sci-nnov.ru}
         \and
             Institute of Astronomy and Astrophysics, Academia Sinica.
P.O. Box 23-141, Taipei 10617, Taiwan, R.O.C.
\and
Ural Federal University, 19 Mira street, 620002 Ekaterinburg, Russia
}
   \date{Received ; accepted }

 
\abstract{}{We aimed at investigating the structure and kinematics of the gaseous disk and outflows around the massive YSO S255 NIRS3 in the S255IR-SMA1 dense clump.
}{Observations of the S255IR region were carried out with ALMA at two epochs in the compact and extended configurations.
}{We serendipitously detected a new, never predicted, bright 
maser line at about 349.1~GHz, which most probably represents the CH$_3$OH $14_{1} - 14_{0}$ A$^{- +}$ transition. The emission covers most of the 6.7~GHz methanol maser emission area of almost 1$^{\prime\prime}$ in size and shows a velocity gradient in the same sense as the disk rotation. No variability was found on the time interval of several months. It is classified as Class~II maser and probably originates in a ring at a distance of several hundreds AU from the central star.
}{}


   \maketitle
%

\section{Introduction}
Observations of cosmic masers are very important for investigations of physical properties and kinematics of associated astronomical objects as well as for better understanding of the maser excitation mechanisms. Many masing transitions have been discovered already, nevertheless new ones, especially unpredicted, are always important for the tasks mentioned above. 
One type of objects where masers are frequently observed and provide valuable information about, are star forming regions, in particular regions of high mass star formation. We 
recently obtained with ALMA a new data set at 0.8~mm toward a bright representative object of this type --- S255IR, located at the distance of
$ 1.78^{+0.12}_{-0.11} $~kpc \citep{Burns16}.

This object has been a target for many investigations. It consists of several dense clumps. The most prominent one, S255IR-SMA1, represents a rotating hot core, probably a disk around a massive ($ \sim 20$~M$_\odot $) young stellar object S255 NIRS3 \citep{Wang11,Zin12,Zin15}.  Recently this YSO attracted an enhanced attention due to the discovery of its 6.7~GHz methanol maser flare in 2015 \citep{Fujisawa15} and subsequent detection of the accretion burst at NIR wavelengths 
\citep{Caratti16,Stecklum16}. 

In our ALMA data we noticed a very bright, apparently non-thermal line toward this object. Here we present the details of the observations and discuss the results.

\section{Observations}
Observations of the S255 IR region were carried out with ALMA in Band 7 at two epochs during Cycle 4 under the project 2015.1.00500.s. The first observing epoch on 2016 April 21, carried out with the 12-meter array in its C36-2/3 configuration having baselines ranging between 15 m and 612 m, has 42 antennas online within a single execution block. The second epoch of observations on 2016 September 9, with the 12-meter array in C40-6 configuration and baselines ranging between 15 and 3143 m, contain back-to-back two execution blocks with 39 antennas online. Four Frequency Division Mode (FDM) spectral windows centered at around 335.4 GHz, 337.3 GHz, 349.0 GHz, and 346.6 GHz, with bandwidths of 1875.0 MHz, 234.4 MHz, 937.5 MHz, and 1875.0 MHz, respectively, were set up for covering continuum and various molecular line features. The spectral feature reported in this paper situates in the spectral window at 349.0~GHz. A total of 3840 spectral channels are employed for resolving the 937.5 MHz bandwidth, resulting in a spectral channel width of 0.244~MHz and a resolution of 0.488~MHz or equivalently 0.42~\kms\ per channel with Hanning-smooth applied.
For observations taken in the first epoch, J0854+2006 and J0750+1231 were used as the bandpass and flux calibrators, respectively. For the second epoch, J0510+1800 was used for both bandpass and flux calibration purposes. In both epochs, J0613+1708 served as the complex gain calibrator.

All data calibration and reduction were carried out with the Common Astronomy Software Applications \citep{McMullin07}. Data were first calibrated through the ALMA manual or pipeline calibration process. We further employed self-calibration for improving the imaging quality and dynamical range. A continuum image was first made using visibilities in the wide low resolution (continuum) spectral window at 335.4 GHz with obvious spectral contamination removed. Two iterations of self-calibration with solution intervals set to 240s and 60s were than performed for deriving refined phase gains. The final continuum image has a 30\% improvement in the dynamic range as compared to the first image. The self-calibration gain solutions were then applied to all spectral windows for further imaging processes.
The resulting images achieve an angular resolution of $ 0{\farcs}10\times 0{\farcs}15 $ with Briggs weighting with a robust parameter of 0.5.

\section{Results}
In the measured spectra toward the SMA1 clump we see a very strong emission line at about 349.1~GHz. Its peak intensity is higher than the intensity of any other line toward this object by more than an order of magnitude. The brightness reaches about 5.9~Jy/beam, which corresponds to about 3900~K, clearly indicating a non-thermal, most probably masing, nature of this line emission. To the best of our knowledge a maser emission at this frequency has never been reported.

An inspection of the major catalogs of spectral lines (Cologne Database for Molecular Spectroscopy, CDMS --- \citealt{Mueller01,Mueller05} and that of the Jet Propulsion Laboratory, JPL --- \citealt{Pickett98}) shows that the most probable identification is the CH$_3$OH $14_{1} - 14_{0}$ A$^{- +}$ line at 349.1070~GHz with its lower level excitation energy of 243.4483~K. The other nearby lines (within $ \pm 5 $~MHz) belong to complex molecules like acetone, vinyl cyanide, aminoacetonitrile, dimethyl ether, which have never been reported to manifest strong maser emission. 
Methanol has many transitions which are masing at certain conditions. However the transition mentioned above has never been observed or predicted to be masing \citep{Sobolev97,Cragg05,Voronkov12}.

\begin{figure*}
\begin{minipage}[b]{0.68\textwidth}
    \includegraphics[width=\textwidth]{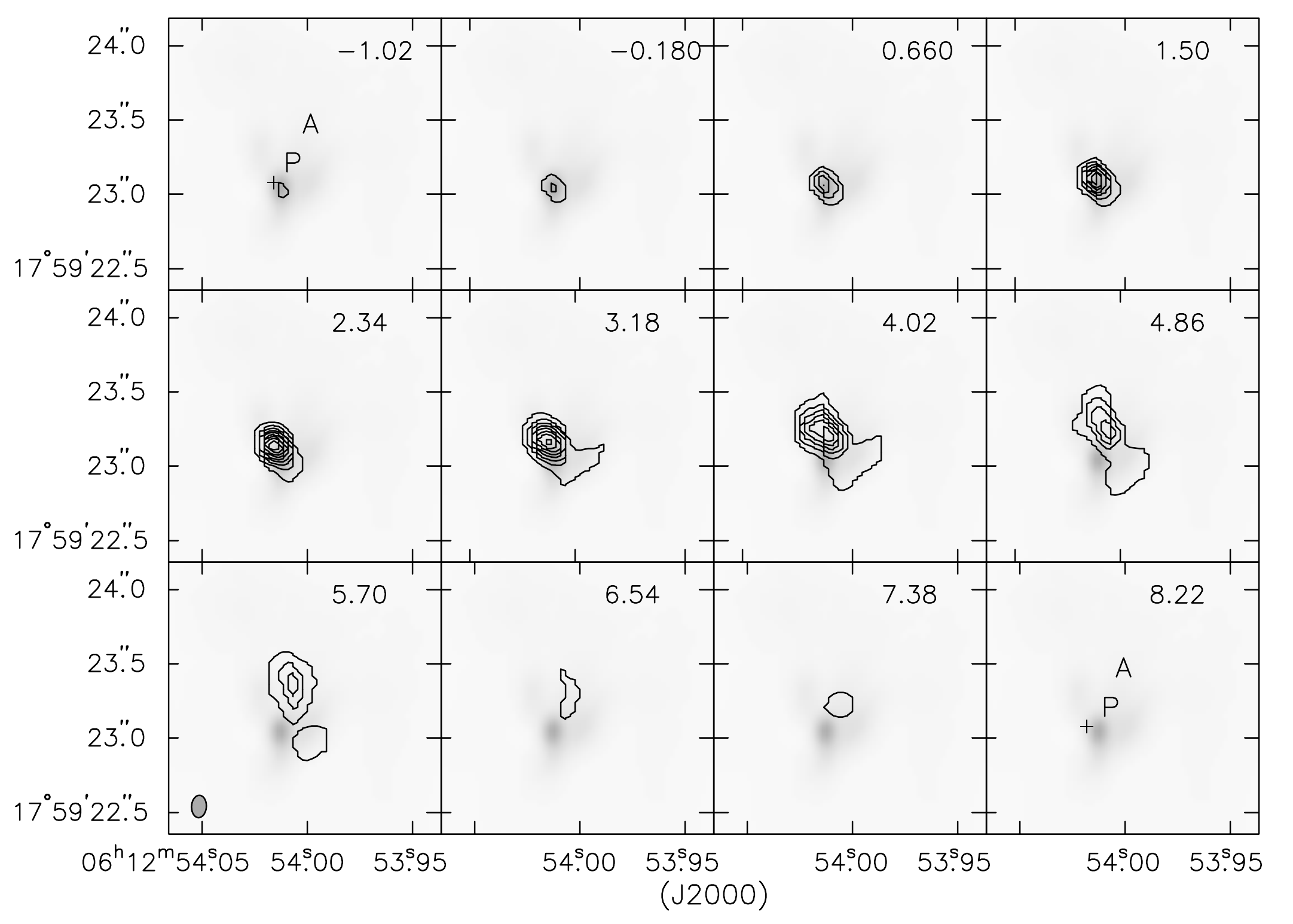}
\end{minipage}
\hfill
\begin{minipage}[b]{0.31\textwidth}
    \caption{The channel map of the CH$_3$OH maser emission toward S255IR-SMA1. The CH$_3$OH emission is shown in contour, with levels at 0.6, 1.2, 1.8, 2.4, 3.0, 3.6, 4.2, 4.8, and 5.4 Jy/beam. The continuum emission at 335~GHz is displayed in greyscale. The datacube has a channel width of 0.42~km/s and maps of only every other channel starting from that at -1.02~\kms are shown. Positions of the 5~GHz radio continuum, and those of the pre-burst and after-burst CH$_3$OH maser clusters reported by \citet{Moscadelli17} are labeled by the cross, P, and A, respectively in the first and last channel for reference. The synthesized ALMA beam is shown in the lower left corner.}
    \label{fig:channel}
\end{minipage}
\end{figure*}

We display in Fig.~\ref{fig:channel} the channel map of this emission (by assuming that the rest frequency equals to that of the CH$_3$OH $14_{1} - 14_{0}$ A$^{- +}$ line). The contours in the map run from 0.6 Jy/beam at the lowest level, corresponding to $\sim$ 400~K, and reach 5.4 Jy/beam, demonstrating the non-thermal nature of the emission. As shown in the figure, strong line emission emerges at about -1~\kms\ spanning to $\sim 8$~\kms\ and its peak location gradually moves toward north-north-west direction.

\begin{figure}
\includegraphics[width=\columnwidth]{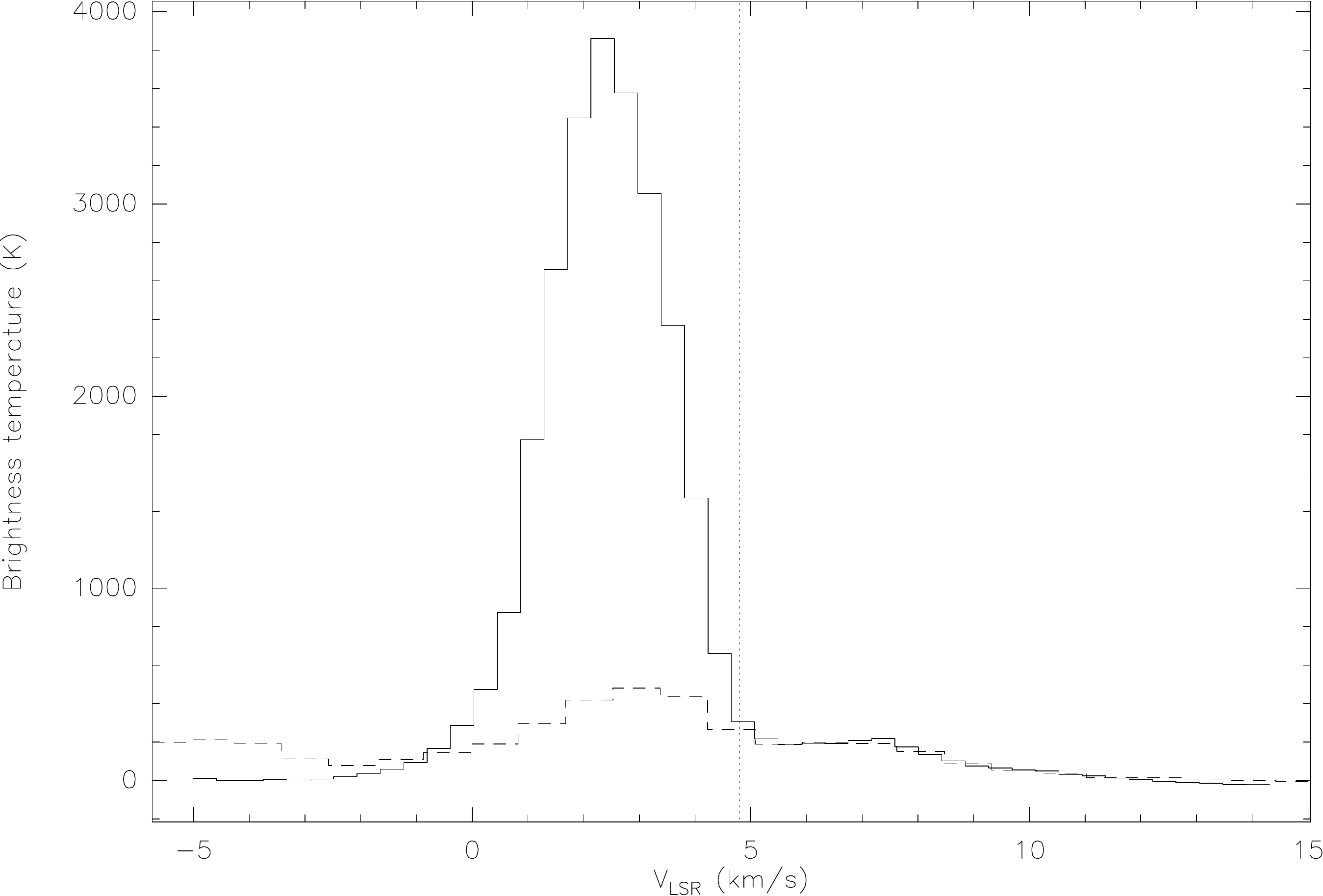} 
\caption{The spectrum of brightness temperature in the masing methanol $14_{1} - 14_{0}$ A$^{-+}$ line on the velocity axis (the zero velocity corresponds to the rest frequency of this transition) within the synthesized beam toward the maser brightness peak (solid line). The dashed line shows the spectrum centered on the CH$_3$OH $16_{1} - 15_{2}$ A$^{-}$ line. The intensity of this line is multiplied by the factor of 3. 
The dotted vertical line indicates the systemic velocity of the S255IR-SMA1 clump (4.8~\kms) as found by \citet{Zin15}. 
}
\label{fig:spec-peak}
\end{figure}

Fig.~\ref{fig:spec-peak} features the spectrum of the maser line toward the brightness peak. For comparison we also show the emission spectrum of one of the apparently thermal methanol lines (the $16_{1} - 15_{2}$ A$^{-}$ line at 345.90397~GHz with the excitation energy of the lower level of 316.0486~K). The profiles of the maser and thermal lines are in general similar {but the blue-shifted part of the maser line is strongly amplified}. The width of the brightest maser feature is about 2.5~\kms.

\begin{figure*}
\begin{minipage}[b]{0.33\textwidth}
\includegraphics[width=\textwidth]{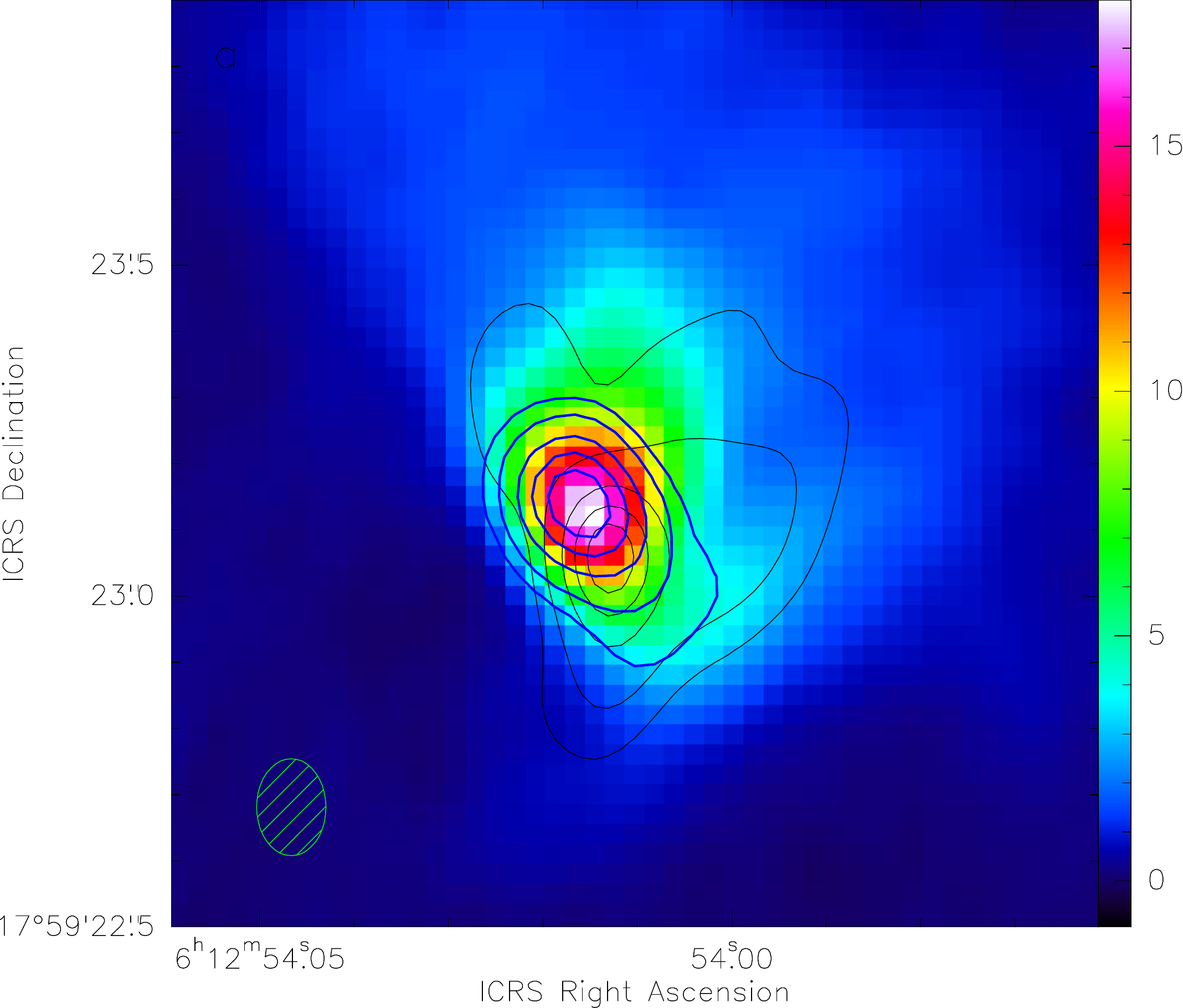} 
\end{minipage}
\hfill
\begin{minipage}[b]{0.33\textwidth}
\includegraphics[width=\textwidth]{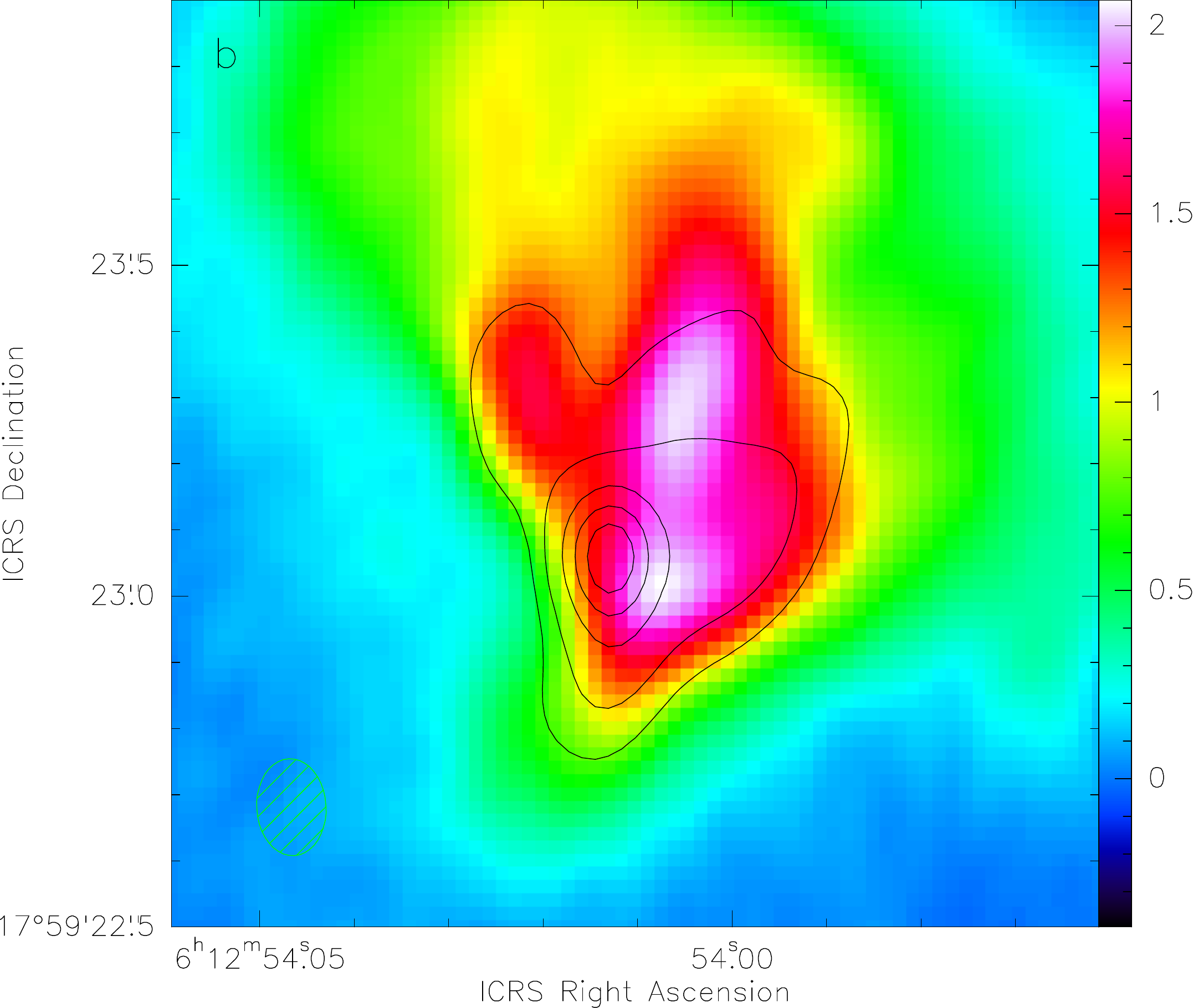} 
\end{minipage}
\hfill
\begin{minipage}[b]{0.325\textwidth}
\includegraphics[width=\textwidth]{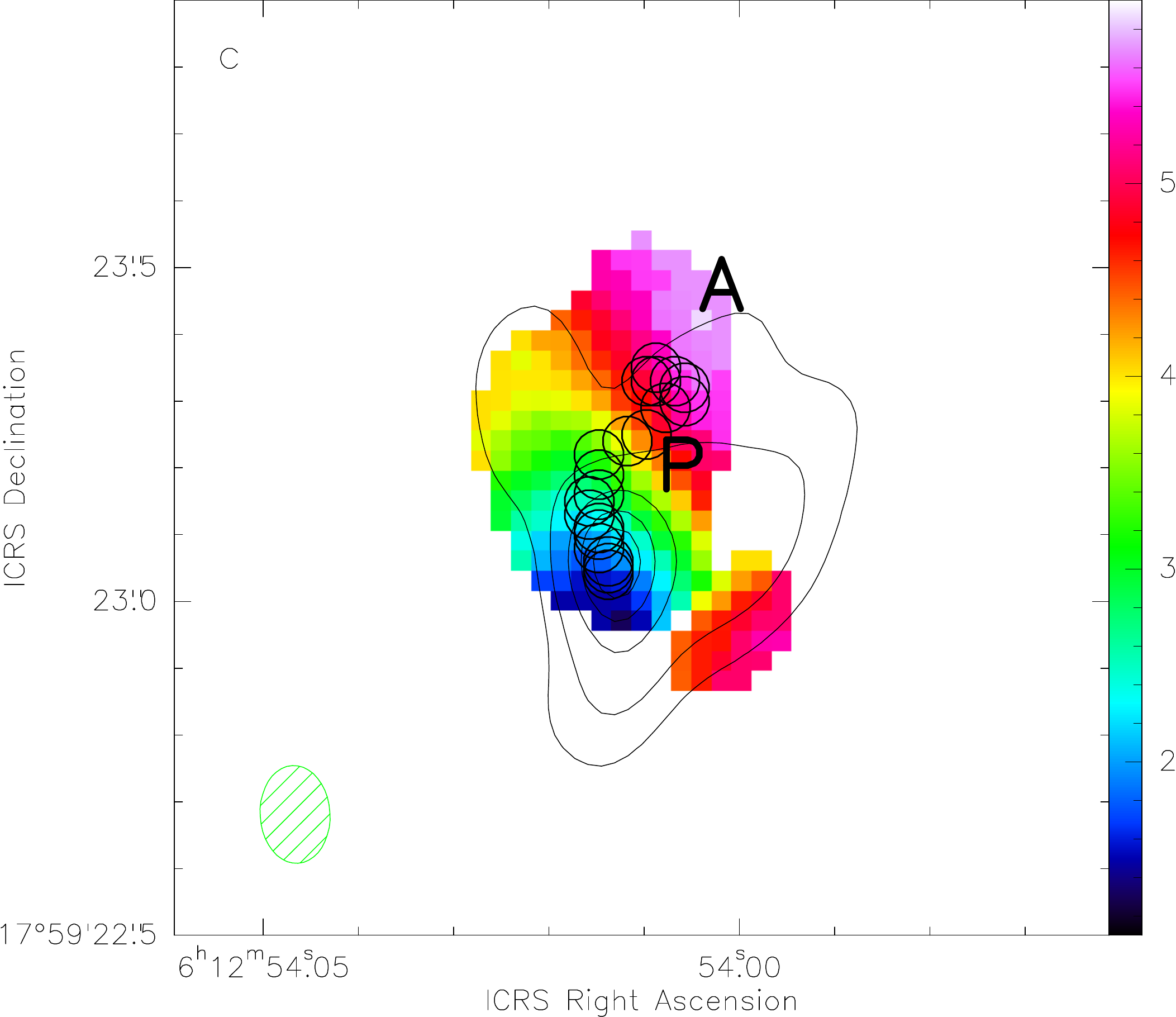} 
\end{minipage}
\caption{Maps of the integrated emission in the CH$_3$OH $14_{1} - 14_{0}$ A$^{- +}$ line (a) and $16_{1} - 15_{2}$ A$^{-}$ line (b). The thick blue contours in the panel ``a" show the map at the velocity of the peak brightness. (c) The 1st moment map of the maser emission with the intensity threshold of 0.9~Jy\,beam$^{-1}$ (which corresponds to approximately 600~K). The chain of circles shows the positions of the emission peaks in different spectral channels. {The locations of the pre-burst and after-burst 6.7~GHz maser clusters are indicated.} The thin contours in all panels show the 0.8~mm continuum emission. The contour levels are 0.1, 0.2, 0.4, 0.6 and 0.8 of the peak intensity. The synthesized ALMA beam is shown in the lower left corner.}
\label{fig:maps}
\end{figure*}

The spatial distribution of the emission in this line is very different from that in the thermal methanol lines, as can be seen in Fig.~\ref{fig:maps} (a) and (b), where we present for comparison also the map of the methanol $16_{1} - 15_{2}$ A$^{-}$ line. The latter follows in general the continuum emission while the former appears relatively centrally peaked with its peak slightly shifted NE from the continuum peak. 
The peak flux density integrated over a circle of about 1$^{\prime\prime}$ in size is about 25~Jy (at velocity of $\sim 4$~\kms).

\citet{Moscadelli17} reported several clusters of the 6.7~GHz maser spots observed with the JVLA and EVN. In Fig.~\ref{fig:channel} the position of their 5~GHz radio continuum as well as those of the dominating pre-/post-burst 6.7~GHz methanol maser clusters labeled as P and A by \citet{Moscadelli17} are marked. The emission detected in our observations covers most of their 6.7~GHz methanol emission area of almost 1$^{\prime\prime}$ in size.
The main peak of our emission, however, is closer to one of their (unlabelled) clusters of maser spots north-east of the 5~GHz continuum.
There is a secondary emission peak, southwest of the main one, which coincides with another unlabelled cluster of the maser spots in \citet{Moscadelli17}.
The emission in individual spectral channels is much more compact. At the velocity of the peak brightness ($\sim 2.34$~\kms) it is about $0\farcs17\times 0\farcs08$ ($\sim 300\times 140$~AU).
There is a significant velocity gradient across the emission region as noted earlier.
This gradient, best demonstrated by the 1st moment map of the emission in Fig.~\ref{fig:maps}c, is in the same sense as the rotation of the disk \citep{Wang11,Zin15}.

\begin{figure}
\centering
\includegraphics[width=\columnwidth]{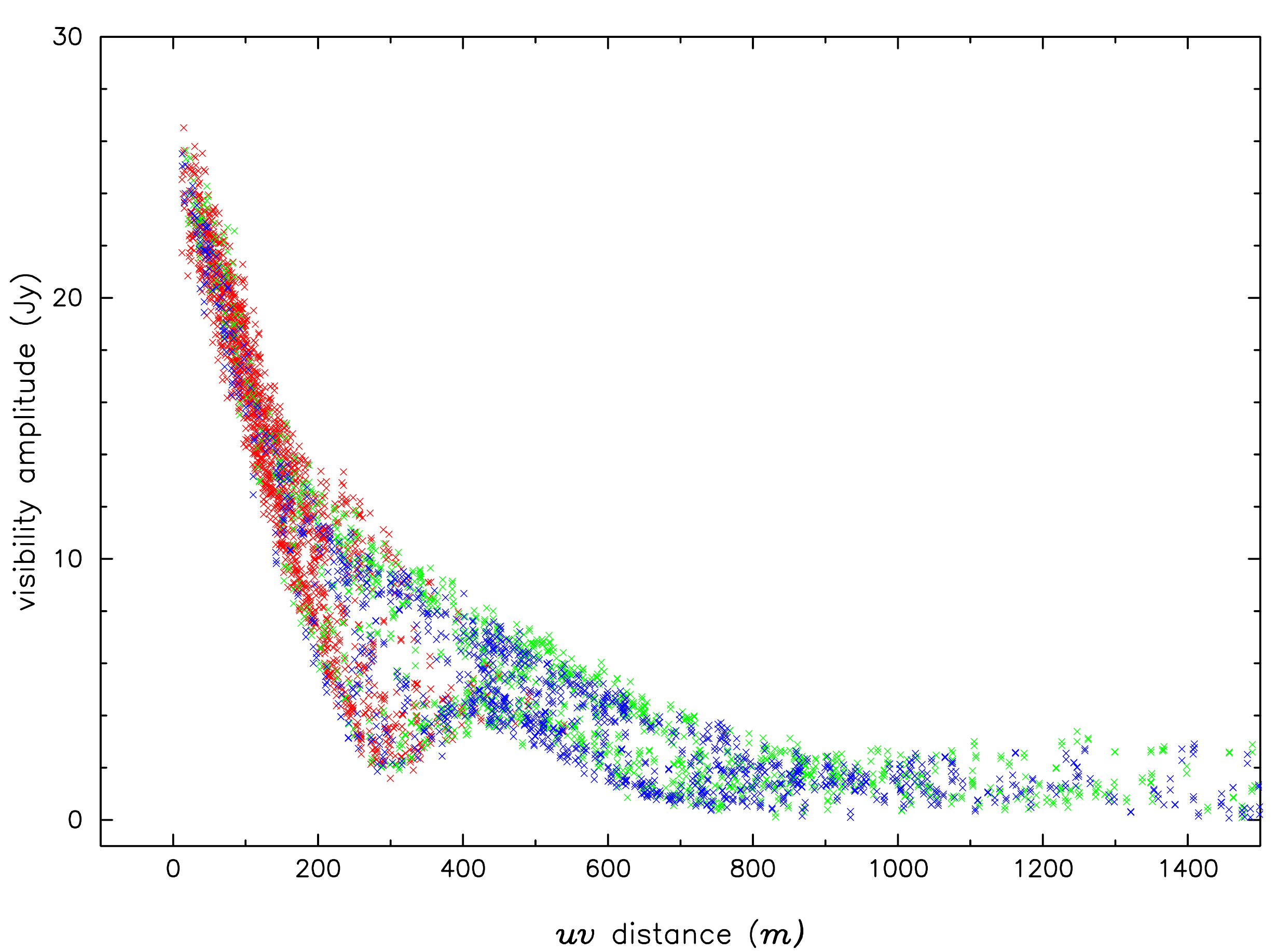} 
\caption{A display of the visibility amplitudes of the maser emission
against the baseline lengths. The visibilities are taken from the
maser emission peak channel at 4~\kms\ from the three execution
blocks (one from 2016 April 21 in red, and two from 2016 September
9 in green and blue). Only visibilities with baseline lengths less
then 1500 m are shown.}
\label{fig:pltuv}
\end{figure}

Maser emission is known to be time variable and we examined
this possibility given that our two epochs of observations are
separated by about five months. 
Nevertheless, the array configurations and hence the angular resolutions in the two observational epochs are different, a direct comparison of the observed flux densities in the image domain through separate imaging of the different individual
observational datasets is not straightforward. 
We instead compare the visibility amplitudes of the maser emission observed at different epochs. 
Figure~\ref{fig:pltuv} displays the visibility amplitudes of the maser emission against the baseline lengths. The visibilities are taken from the integrated emission peak channel at around 4~\kms\ from the three execution blocks (one from 2016 April 21 in red, and two from 2016 September 9 in green and blue). 
As shown in the figure, the visibilities from the three datasets have fully consistent amplitude distribution within baseline lengths shorter than about 180 meter, which corresponds
to 205~k$\lambda$, or an angular scale of $\sim 1^{\prime\prime}$. 
This implies no noticeable flux density variation at an angular scale $\geq 1^{\prime\prime}$. The visibility amplitudes at intermediate baseline lengths (between 180~m and $\sim 600$~m) appear to have primarily two branches of amplitudes with those from 2016 April mainly in the higher branch and those from September in the lower branch. We verify that it is because these visibilities from the two epochs are sampling different parts (direction) of the $uv$ domain. Still, the amplitudes scales of the overlapping region are fully consistent.
The consistency of amplitudes between visibilities from different epochs also preserve throughout the methanol emission channels, suggesting no line profile variation between the observations. We have no sampling at long baselines in April to differentiate or identify any flux density variation at angular scales significantly less than $0{\farcs}4$. Given that the absolute flux calibration for band 7 is at a level of about 5\% based on the ALMA Cycle 4 Technical Handbook, we conclude that the methanol emission shows no variation above this
level in its total flux density over a timescale of several months.

\section{Discussion}
Analysis of observational data had shown that there are two classes of methanol masers, Class~I and Class~II, which emit in the different sets of transitions \citep{Menten91}. Masers of one class never display strong emission in transitions characteristic of the other class.
Observational results are supported by theoretical modelling which shows that the warm radiation field cools down collisionally excited maser transitions and provides power to transitions which are suppressed by the pumping mechanism with collisional excitation \citep{Sobolev07,Voronkov05, Cragg92}. As a result, Class II methanol masers reside closer to YSOs and trace circumstellar disks and inner parts of the outflows while Class I masers are associated with more distant parts of the outflows and shocked regions. This conclusion finds observational support in, for example, \cite{Sanna15,Sanna17} and \cite{Voronkov14}.

Though methanol masers of different classes are formed under different conditions they can be located close to each other or even coincide due to projection effects \citep[e.g.][]{Kurtz04,Ellingsen05,Sobolev07}. So, knowledge about the class of maser transition is important for understanding which parts of the YSO vicinity are traced. 

Existence of two classes of methanol masers can be qualitatively explained by selection rules for radiative transitions which lead to overpopulation of the sets of energy levels with particular value of quantum number $K$ (called $K$-ladders) with respect to the levels from the different $K$-ladder \citep{Menten91}. As a result maser transitions form series of transitions which differ by the value of quantum number $J$. The 349.1070~GHz transition under study belongs to $J_{1} - J_{0}$ A$^{- +}$ series. Maser emission in transitions of this series was not observed or theoretically predicted. However, qualitative considerations show that this transition probably belongs to Class~II.
Observations show that the strongest Class~II maser occurs in $5_{1}~-~6_{0}$~A$^{+}$ transition while the strongest Class~I maser occurs in $7_{0}~-~6_{1}$~A$^{+}$ transition. So, Class~II methanol masers are characterized by overpopulation of A$^{+}$ levels of $K=1$ ladder with respect to the A$^{+}$ levels of $K=0$ ladder while for Class~I masers the situation is opposite. Model calculations for Class~II methanol maser conditions by \cite{Cragg05} show that A$^{-}$ levels of $K=1$ ladder tend to be overpopulated with respect to A$^{+}$ levels of $K=1$ ladder. 
This tells that the 349.1070~GHz transition from $J_{1} - J_{0}$ A$^{- +}$ series most probably belongs to Class~II family.
Strong support to this conclusion comes from the observational evidence that emission in the newly discovered maser transition covers the same velocity range and shows similar position-velocity dependence as the strongest Class~II methanol maser at 6.7~GHz \citep{Moscadelli17}. {The spectra and the peak locations of the new maser may not coincide with those of the 6.7~GHz ones due to  different sensitivity to the physical conditions which is usual for Class~II methanol maser transitions \citep{Cragg05}.} 

The $J_{1} - J_{0}$ A$^{- +}$ line series contains transitions with the frequencies starting from 303.37~GHz. Many of them can be observed with the same instrument, which allows to avoid calibration problems in determining the relative line intensities. The lower levels of these transitions cover rather wide range of energies from 2.32~K for the $1_{0}$~A$^{+}$ level up to 243.45~K for the $14_{0}$~A$^{+}$ lower level of the transition under study and higher. It is worth mentioning that in our SMA data obtained in December 2010, the CH$_3$OH $13_{1} - 13_{0}$ A$^{- +}$ line at 342.729796~GHz was detected in quasi-thermal emission \citep{Zin15}. {This may indicate that the discovered maser is related to the accretion burst in 2015, however additional modeling and monitoring of these lines are needed for exact answer.} Observations of the methanol line series proved to be a good tool for the physical parameter diagnostics both in maser and quasi-thermal cases |\citep{Voronkov06,Menten88,Salii06}. This is very important because the new maser line was detected toward positions spread over the disk. 

An inspection of our channel maps shows that the peaks of the maser emission at different velocities lie on the curve, which resembles an arc (Fig.~\ref{fig:maps}c). This can be a part of a ring in the disk or on the surface of the disk at a substantial distance of several hundreds AU from the star. Taking into account the disk orientation, which follows from the outflow data \citep{Zin15,Burns16}, it becomes clear that the brightest emission comes from the closer to us part of the disk. For the lower brightness component the velocity gradient is less clear but its location shows that it may originate on the far side of the disk. The difference in brightness may be related to the different background level.



\section{Conclusions}

We report the discovery of a new, never predicted rather bright maser line at about 349.1~GHz. Most probably this line represents the CH$_3$OH $14_{1} - 14_{0}$ A$^{- +}$ transition with the excitation energy of the lower level of 243.4483~K. It can be classified as a Class~II methanol maser.

This line was detected toward the S255IR-SMA1 clump, which represents a rotating disk around the massive YSO S255 NIRS3. The maser emission is observed in the same area as the 6.7~GHz methanol maser and shows a velocity gradient in the same sense as the disk rotation. It probably originates in a ring at a distance of several hundreds AU from the central star. No variability was found on the time interval of several months.

\begin{acknowledgements}
We are grateful to the anonymous referee for the helpful comments.
IZ was supported by the Russian Science Foundation (project No. 17-12-01256).
This paper makes use of the following ALMA data: ADS/JAO.ALMA\#2015.1.00500.S. ALMA is a partnership of ESO (representing its member states), NSF (USA) and NINS (Japan), together with NRC (Canada), MOST and ASIAA (Taiwan), and KASI (Republic of Korea), in cooperation with the Republic of Chile. The Joint ALMA Observatory is operated by ESO, AUI/NRAO and NAOJ.
\end{acknowledgements}

\bibliographystyle{aa} 
\bibliography{maser} 

\end{document}